\documentclass[natbib]{svjour3}                     
\smartqed  
\usepackage{graphicx}
\newcommand{\ch}{{\it Chandra}}
\newcommand{\xmm}{XMM-{\it Newton}}
\newcommand{\mrk}{Mrk~279}

\newcommand{\Lya}{\ifmmode {\rm Ly}\alpha \else Ly$\alpha$\fi}
\newcommand{\Lyb}{\ifmmode {\rm Ly}\beta \else Ly$\beta$\fi}
\newcommand{\Lyg}{\ifmmode {\rm Ly}\beta \else Ly$\gamma$\fi}

\newcommand{\cii}{C\,{\sc ii}}
\newcommand{\ciii}{\ifmmode {\rm C}\,{\sc iii} \else C\,{\sc iii}\fi}
\newcommand{\civ}{\ifmmode {\rm C}\,{\sc iv} \else C\,{\sc iv}\fi}
\newcommand{\cv}{\ifmmode {\rm C}\,{\sc v} \else C\,{\sc v}\fi}
\newcommand{\cvi}{\ifmmode {\rm C}\,{\sc vi} \else C\,{\sc vi}\fi}

\newcommand{\nv}{N\,{\sc v}}

\newcommand{\ov}{O\,{\sc v}}
\newcommand{\ovi}{O\,{\sc vi}}
\newcommand{\ovii}{O\,{\sc vii}}
\newcommand{\oviii}{O\,{\sc viii}}

\newcommand{\neix}{Ne\,{\sc ix}}

\newcommand{\mgii}{Mg\,{\sc ii}}

\newcommand{\siIII}{Si\,{\sc iii}}

\newcommand{\ferii}{Fe\,{\sc ii}}

\newcommand{\fexxv}{Fe\,{\sc xxv}}
\newcommand{\fexxvi}{Fe\,{\sc xxvi}}

\newcommand{\kms}{km\,s$^{-1}$}

\newcommand{\aap}{{Astron. \& Astrophys.}}
\newcommand{\apj}{{Astrophys. J.}}
\newcommand{\apjs}{{Astrophys. J. Supp. Ser.}}
\newcommand{\apjl}{{Astrophys. J. Letters}}
\newcommand{\mnras}{{Monthly Notices of the Royal Astronomical Society}}

\newcommand{\araa}{{ARA\&A}}

\newcommand{\pasp}{{Pub. of the Astr. Society of the Pacific}}
\begin{document}

\title{The Ultraviolet-X-ray connection in AGN outflows}

\titlerunning{The UV-X-ray connection}        

\author{Elisa Costantini         
}


\institute{E. Costantini \at
              SRON, Netherlands Institute for Space Research, Sorbonnelaan 2, 3584CA, Utrecht, The Netherlands \\
              \email{e.costantini@sron.nl}           
}

\date{Received: date / Accepted: date}

\maketitle

\begin{abstract}
In this paper I review the recent progress in understanding the physics of the gas outflowing from active galactic
nuclei and its impact on the surrounding environment, using the combined information provided by multiwavelength Ultraviolet-X-ray campaigns.  

\end{abstract}

\section{Introduction}
\label{par:intro}

Theoretical studies predict that the mass ejected by active galactic nuclei (AGN) has strong implications 
on the AGN self-sustenance \citep[e.g.][]{2008ApJS..175..356H} and its contribution to the environment 
\citep[e.g.][]{2009ApJ...699...89C,2010ApJ...717..708C}. 
In particular, the outflowing gas can have an important role in host galaxy enrichment 
\citep[][and references therein]{2008MNRAS.391..481S}, up to larger scale environments
like the intra-cluster medium \citep[][for a review]{2007ARA&A..45..117M} and eventually the intergalactic medium 
\citep[e.g.][]{2004ApJ...608...62S,2007ApJ...665..187L}. 
Metal enrichment of the environment is likely to act both at low- and high-redshift 
 \citep{2005ApJ...635L..13S,2010ApJ...721..174M}. As a consequence, an impact 
 on the cosmological quasar luminosity function is expected \citep[e.g.][]{2006ApJ...650...42L,2007ApJ...654..731H}. 
Ionized-gas outflows (also called warm absorbers) are detected in roughly 50\%
of the AGN X-ray spectra. There is a one-to-
one correspondence between objects that show
X-ray absorption and those that show intrinsic
UV absorption, indicating that these two phenomena are related \citep{1999ApJ...516..750C}. Multiwavelength studies
of this AGN feature represent a unique tool to globally study the physics of this gas \citep[][]{2003ARA&A..41..117C}.

\section{Early studies}
\label{par:early}

Active Galactic Nuclei (AGN) have been know to host ionized gas first through their absorption and emission features in
their optical spectra \citep[][]{1943ApJ....97...28S}.
In the UV band, narrow
absorption lines from low- (e.g. \mgii) and high-ionization (e.g. \civ, \nv ) atoms have been frequently detected by UV
satellites in the 80's, \citep[e.g. IUE;][]{1983ApJ...267..515U}. At the same time, many of the sources exhibiting UV
absorption, showed a low-energy cut-off in the X-rays \citep[EXOSAT;][]{1989MNRAS.240..833T}. Early attempts of depicting
a common origin for the UV and X-ray gas, were not successful \citep{1988MNRAS.230..121U}. The low resolution of early
instruments played the major role in confusing the physical picture, as, for instance, 
ionized absorption could be hardly distinguished from neutral absorption. The advent of medium-resolution X-ray
instruments allowed a major step forward, thanks to the detection of the \ovii\ and \oviii\ 
\citep[e.g.][]{1998ApJS..114...73G}. Albeit absorption features suffered from blending, e.g. \ovii\ was blended with the iron
unresolved transition array (UTA), a column density and a degree of 
ionization (Sect.~\ref{par:ion_par}) could be assigned to the gas absorbing the X-rays. Several studies could then put in relation
the X-ray gas and the more highly ionized species (\nv, \civ, \Lya) detected in the UV
\citep[e.g.][]{1994ApJ...434..493M,1995ApJ...452..230M}.

\begin{table}[t]
\caption{Recent multiwavelength campaigns on bright AGN. }
\label{t:list}       
\begin{tabular}{llll}
\hline\noalign{\smallskip}
Object &  year & Instruments  & Ref.  \\
\noalign{\smallskip}\hline\noalign{\smallskip}
NGC~3516 & 2000 & LETG, STIS & 1,2\\
NGC~3783 & 2000& HETG, STIS, FUSE& 3,4\\
NGC~4051 & 2000 & HETG, STIS &5 \\
Mrk~509 & 2001 & HETG, STIS& 6,7\\
 & 2009 & RGS, LETG, COS& 8\\
NGC~4151 & 2002 & HETG, STIS &9,10\\
NGC~7469 & 2002 & HETG, STIS, FUSE&  11\\
Mrk~279 &  2002 & HETG, STIS, FUSE& 12\\
 &  2003 & LETG, STIS, FUSE& 13,14\\
NGC~5548 & 2003 & HETG, STIS & 15,16\\
1H~0419-577 & 2010 & RGS, COS& 17\\
\noalign{\smallskip}\hline
\end{tabular}

\noindent
References: 
(1) \citet{2002ApJ...571..256N}, 
(2) \citet{2002ApJ...577...98K}, 
(3) \citet{2002ApJ...574..643K}, 
(4) \citet{2003ApJ...583..178G}, 
(5) \citet{2001ApJ...557....2C}, 
(6) \citet{2003ApJ...582..105Y}, 
(7) \citet{2003ApJ...582..125K}, 
(8) Kaastra et al. in prep., 
(9) \citet{2005ApJ...633..693K}, 
(10) \citet{2006ApJS..167..161K}, 
(11) \citet{2005ApJ...634..193S},
(12) \citet{2004ApJS..152....1S}, 
(13) \citet{2005ApJ...623...85G}, 
(14) \citet{2007A&A...461..121C}, 
(15) \citet{2003ApJ...594..116C}, 
(16) \citet{2005A&A...434..569S}, 
(17) Costantini et al. in prep.

\end{table}

\section{Methods}
\label{par:methods}
The increase of energy resolution and sensitivity of the instruments in both UV (HST-STIS, HST-COS, FUSE) and X-ray band
(\ch-LETGS, \ch-HETGS and \xmm-RGS) revealed the great
potential of a multiwavelength studies of this gas. This approach does not merely offer a broader-band view of the
phenomenon, but allows to exploit the complementarity of the instruments to infer physical quantities. Essential
ingredients of these campaigns are of course the simultaneous pointing of the AGN and an adequate exposure time, in order to
maximize the signal-to-noise ratio. In Table~\ref{t:list} recent multiwavelength campaigns are listed, along with representative references for the
UV and X-ray analysis. Those delivered not
only information on individual sources, but also provide mean properties of the gas in this class of objects.\\
The higher spectral resolution provided in the UV band (R$\sim$20,000) allows a detailed, velocity dependent, study of the
absorption troughs of a few important ions (e.g. \civ, \nv, the Lyman series, \ovi). The X-rays provide a blurred (i.e. lower
resolution R=400-1000) view of dozens of transitions of He- and H-like ions of C, N, O, Fe, Ne, Si and S. In the
following I review in which way the information from the two bands can be combined.

\subsection{Kinematics}
Multiwavelength campaigns show that the UV absorber is composed by as much as four or more components with distinct
outflow velocities, ranging from few tens to thousands \kms \citep[e.g.][]{2003ApJ...583..178G}. The outflow velocity is
measured as the displacement of the centroid of the absorption line with respect to its wavelength in 
the rest frame of the source, converted to velocity space. The X-ray absorbing gas usually shares some of the
UV velocity components. The lower X-ray resolution however allows to assign a range of possible velocities which can be
usually associated to multiple UV components with similar velocity 
\citep[e.g.][]{2002ApJ...574..643K,2005A&A...434..569S}. 
However, a higher velocity component, with no correspondence in the UV band, is sometimes detected in the X-ray band only 
\citep[e.g.][]{2001ApJ...557....2C}. This gas component is associated with the higher ionization ions, like \oviii,
suggestive of an additional, highly ionized, gas in the system. However, no systematic correlation between outflow
velocity and ionization parameter has been found in the X-rays band \citep{2005A&A...431..111B}. 

\subsection{Ionization parameter}\label{par:ion_par}
An ionized gas is phenomenologically 
parameterized not only by its kinematic components, but also by its ionization and column density. 
Two similar definitions of the ionization parameter are used. The dimentionless parameter $U$ is expressed as 
\begin{math}
U=\int^{\infty}_{\nu0}\frac{L_{\nu}/h\nu}{4\pi r^2n_{\rm H}c}d\nu 
\end{math}
where $L_{\nu}/h\nu$ is the rate of emitted photons per frequency, $n_{\rm H}$ the hydrogen number density and $r$ the gas distance. The luminosity is integrated over the
total ionizing spectrum, from $\nu_0$, the ionization threshold of hydrogen. The other commonly used parameter is
$\xi=L_{ion}/n_{\rm H}r^2$ (erg\,cm\,s$^{-1}$), where $L_{ion}$ is the integrated luminosity. 
Sometimes, especially in X-ray studies, $\nu_0$ is taken as 0.1 keV \citep[e.g.][]{1998ApJS..114...73G}. Here we will refer to the ionizing parameter as the log of 
$\xi$.

In order to find a solution for $\xi$, a ionization balance is needed. 
This expresses the balance between the recombination to and the ionization of a
given ion and it is dependent on the spectral energy distribution (SED) of the source. The
ionization balance is calculated using a photoionization code, such as Cloudy \citep{1998PASP..110..761F} 
or XSTAR \citep{2001ApJS..133..221K}. A grid is constructed for a range of values of $\xi$ and N$_{\rm H}$ and the best
fit searched through a minimizing procedure like the $\chi^2$ or the $C^2$ statistical tests.
In the UV band, this procedure relies on the analysis relatively few ions (Sect.~\ref{par:methods}). 
Despite the lower resolution, a solution for 
$\xi$ and N$_{\rm H}$ can be easily found using X-ray data, thanks to the dozens of transitions available. In particular the presence of
the iron UTA, where virtually all Fe ions are represented, helps in securing the parameter determination.

\begin{figure*}
\begin{center}
  \includegraphics[angle=90,width=0.6\textwidth]{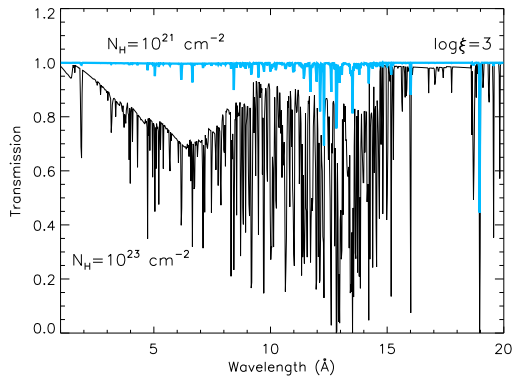}
\end{center}
\caption{Example of two highly ionized gas (log$\xi=3$), with differing column density. The resolution assumed is the one
of the \ch-HEG grating. It is evident that a high column density is necessary to clearly highlight the absorption
features in an ordinary, relatively low s/n spectrum.}
\label{f:high:xi}       
\end{figure*}

Higher ionization ions can be detected only in the X-ray band. Highly ionization components are highlighted mostly 
by H-like and He-like ions of O, Ne, Mg, Si, S and also Fe, detected at very short wavelength ($\lambda$=1.85 and 1.79
for \fexxv\ and \fexxvi, respectively). Fig.~\ref{f:high:xi} shows that the opacity of such gas is modest, unless the
column density becomes high ($N_{\rm H}>$ few$\times10^{22}$\,cm$^{-2}$). This observational bias may well be the reason
why the higher ionization gas often displays a higher column density \citep[e.g.][]{2008A&A...483..161T}. There could be
also low $N_{\rm H}$ gas which eluded the detection so far. It is evident however that a larger hydrogen column density
is associated with 
the detected high ionization outflowing gas. If the UV and X-ray absorbers are co-spacial,
this larger column density implies that a larger mass is associated with the X-ray absorber than with the UV absorber. 
Therefore the X-ray gas may play an important role in the 
budget of material carried away from the black hole.
\subsection{Column density}\label{par:nh}

\begin{figure*}
\begin{center}
  \includegraphics[width=0.5\textwidth]{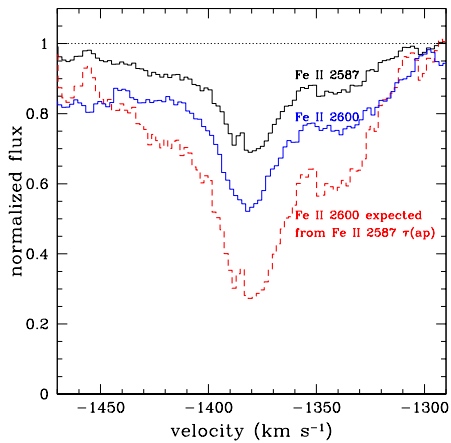}
\end{center}
\caption{Normalized residual
intensity for \ferii\ 2587 and \ferii\ 2600\,\AA\ are shown. 
The predicted \ferii\ 2600\,\AA\ computed from the optical depth
ratio (lower curve) fails to reproduce the observed (middle curve) data \citep{2008ApJ...681..954A}.}
\label{f:tau}       
\end{figure*}

In general there is a one-to-one relation between objects that show X-ray absorption and those which show UV absorption. 
In particular, these multiwavelength studies (Table~\ref{t:list}) clearly pointed out that 
at least one UV gas component shares the same outflow velocity and ionization parameter with one X-ray component.\\ One
major problem that confused for a while the interpretation of this phenomenon is the non-correspondence of column density between the
gas detected at different frequencies. However, detailed studies revealed that the two values 
can be most of the times reconciled if a velocity-dependent partial covering, acting in the UV band, is allowed
\citep[e.g.][]{2002ApJ...566..699A}. 
The integrated column density of an absorption trough is 
\begin{math}  
N\propto 1/\lambda f\int \tau({\sl v})d{\sl v},  
\end{math} 
where $\tau$ is the optical depth, $\lambda$ is the wavelength of the transition,  $f$ is the oscillator strength and
{\em v} is the velocity. 
The residual intensity of a line is given by $I_1=exp(\tau_1)$. If the line is part of the doublet (or multiplet), the
optical depth ratio of two lines at ground level will be $R_{21}=f_2\lambda_2/f_1\lambda_1$. As a consequence, the
residual intensity of the other term of the doublet can be also predicted: $I_2=exp(-R_{21}\tau_1)=I_1^{R_{21}}$.
However, observations show a very different behavior, which is illustrated in Fig.~\ref{f:tau}. The observed blue part
of the line multiplet is significantly shallower than predicted, implying that a covering factor 
$C_f(v)\simeq1-I_2(v)$ is at work. Accounting for $C_f(v)$, a column density compatible with the X-ray gas
is obtained in most cases. The corrected column density allowed also a correct estimate of element abundances 
\citep[see, e.g.][]{2001ApJ...548..609D}. In at least one case, even absolute abundances could be evaluated 
\citep{2007ApJ...658..829A}. X-ray lines do not suffer dramatically from the covering factor effect in
these objects, 
probably because the X-ray
source is more compact than the UV continuum, thus the gas can more easily cover our line of sight. 
In one noticeable case, where the 
3rd order of HETG data could be analyzed \citep[NGC~3783,][]{2002ApJ...574..643K}, the striking similarities between the profile of 
\Lyb, detected in the UV, and the one of \ovii, detected in the X-rays, suggests however that a covering factor could be important
also for the X-ray gas \citep[see Fig.~2 of ][]{2003ApJ...583..178G}. For a typical data set, higher dispersion 
order data, implying an enhanced energy resolution, are of poor quality. This makes impossible in practice 
the detection and the analysis of a velocity-dependent covering factor. \\

\begin{figure}[t]
\begin{center}
\hbox{
\begin{minipage}{5.5cm}
\includegraphics[width=5.5cm,height=7cm]{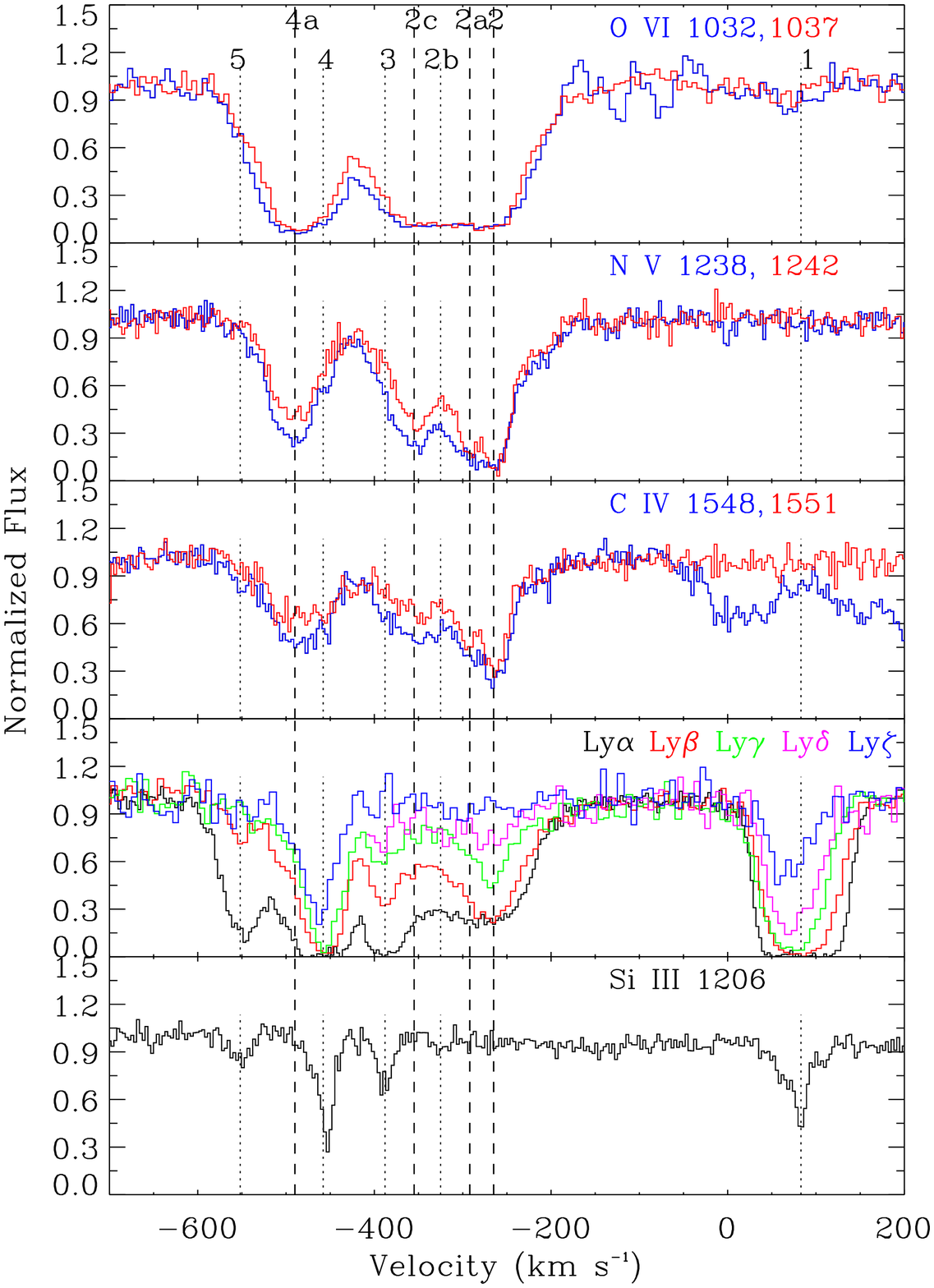}
\end{minipage}
\hspace{0.3cm}
\begin{minipage}{6.5cm}
\includegraphics[width=6.5cm,height=8.5cm]{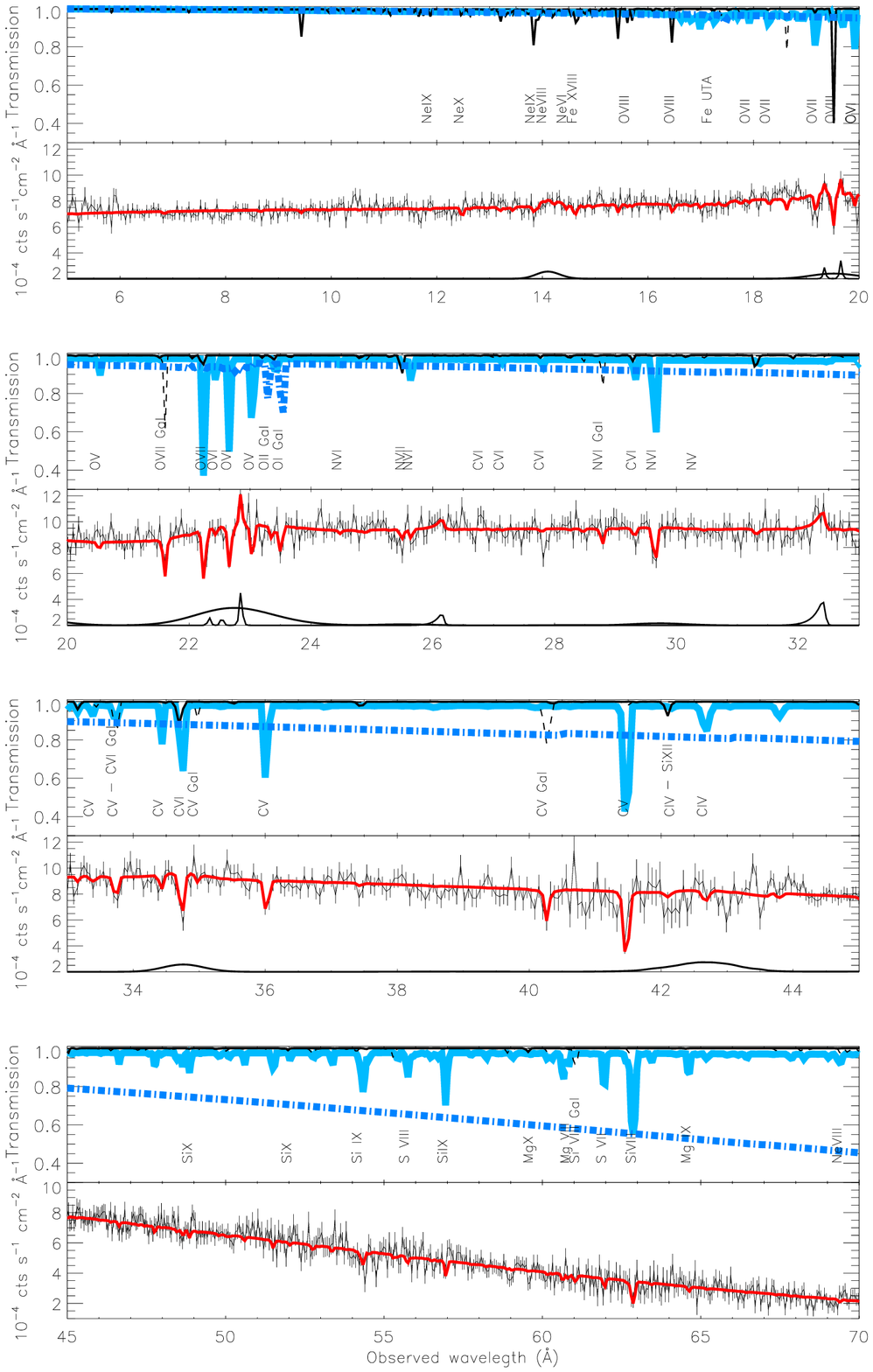}
\end{minipage}
}
\end{center}
\vspace{0.5cm}
\caption{(Left panel): Normalized absorption profiles from the 
STIS and FUSE spectra of \mrk. The spectra are plotted as a function of velocity with respect to the systemic
redshift of the host galaxy. The dashed vertical lines identify the velocity components 
\citep{2005ApJ...623...85G}. (Right panel): X-ray spectrum as measured by \ch-LETGS along with the best fit model
and the separate emission components (black lower line). 
In the upper panels the transmission of the gas components are shown \citep{2007A&A...461..121C}.}  
\label{f:mrk279}       
\end{figure}

The phenomenological picture that emerges from the comparison of the absorption features in the two wave bands is that
lower ionization warm absorbers are naturally detected in the UV. These are characterized by e.g. \mgii, \cii. Then
there is a phase which is detected in both bands. This is characterized e.g. by \nv, \civ, \ovi, \ovii. Note that \ovi\
is the only abundant ion which is simultaneously detected in the FUV and X-rays. Unfortunately \ovi, which was
successfully studied using FUSE in nearby objects (Table~\ref{t:list}), 
is out of the reach of the high-resolution instruments now flying on board of HST\footnote{The \ovi\ line enters the
 COS spectrograph wavelength band for redshift z$>0.1$.}. 
Finally, there is a high ionization phase (log$\xi>2$) which can be only
detected in the X-rays.\\
A typical example is given in Fig.~\ref{f:mrk279} where both the UV and X-ray 
representative spectra of \mrk\ are shown. The UV spectrum displays many velocity components of which only component $2$ is shared
with the X-rays. For that component, a full agreement of column density and ionization parameter is found in the two
spectral bands \citep{2007ApJ...658..829A}. Lower ionization ions (e.g. \siIII, Fig.~\ref{f:mrk279}) 
are represented in the UV spectrum only \citep{2005ApJ...623...85G}. On the
other hand, the high-ionization component, highlighted mainly by \oviii\ (correspondent to a log$\xi\sim2.5$, black solid line in the upper right panels
of Fig.~\ref{f:mrk279}) is only detected in the X-ray band \citep{2007A&A...461..121C}.

\section{The origin of the UV-X-ray warm absorber}
In the following I describe what are the important parameters to understand the geometry and distance of the warm
absorber. 

\subsection{The density and distance of the gas}
\begin{figure*}
\begin{center}
  \includegraphics[width=0.6\textwidth,angle=-90]{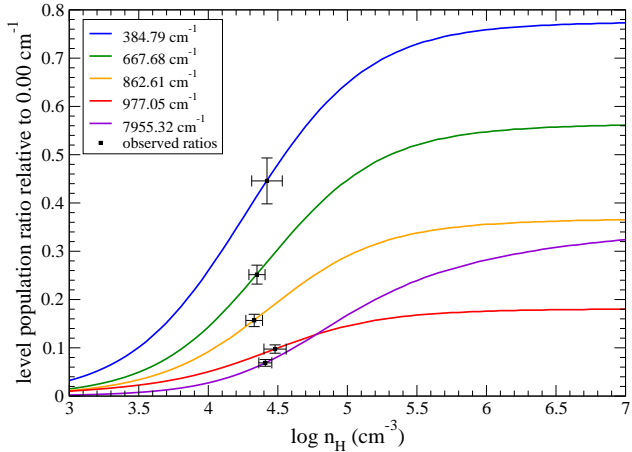}
\end{center}
\caption{Calculated level populations of excited levels of \ferii\ relative to the ground level as a function of the
 the number density. The points
indicate the observed column density ratio with the ground level 
\citep[in this example: QSO 2359-1241,][]{2008ApJ...688..108K}.}
\label{f:n_levels}       
\end{figure*}

One of the quest in the study of the warm absorbers is the determination of the hydrogen number density of the gas. Indeed, 
in both energy bands there is not any straightforward method to obtain this parameter. In the UV, the most promising and
accurate method is the analysis of meta-stable levels. 
These  atomic levels are those from which the transition to the ground level is forbidden. They
are populated by collisions, if the number density is above a certain threshold whose value is in turn dependent on the
transition. Such experiment has been carried out e.g. using \ciii$^{*}$ and \ferii$^{*}$. 
In Fig.~\ref{f:n_levels} the practical procedure is illustrated. For each
level, the theoretical relative population of the meta-stable level with respect to the ground 
level is plotted as a
function of the number density $n_{\rm H}$. This curve is confronted with each observed relative 
population. It is evident that 
the more transitions are available, the more constrained the value of
$n_{\rm H}$ will be \citep[e.g.][]{2005ApJ...631..741G}.

In the X-rays, very few Be-like ions
(as \ciii$^{*}$ is) can be detected. One of them is \ov$^{*}$, whose main transition from the meta-stable level 
lies at $~22.5$\AA. This meta-stable level is sensitive only to high densities, therefore it may be not ubiquitous in AGN spectra. A tentative detection has been reported by 
\citet{2004A&A...428...57K}, implying a number density $n_{\rm H}>3\times10^{12}$\,cm$^{-3}$.\\
Meta-stable levels are sensitive to only a certain range of number densities, depending on the transitions. In principle other
methods, like monitoring the variability of the warm absorber in response to the central source variation, would be 
sensitive to any number density value, provided that the monitoring is carried out on a reasonably extended time base line. 
If one assumes that recombination is the only mechanism at work when the ionized gas responds to a variation in the
ionizing flux, then the number density of the gas is simply inversely proportional to the recombination time 
\citep[e.g.][]{1995ApJ...447..512K,2000ApJ...537..134B}. 
In particular:\\
\begin{math}
\tau_{\mathrm{rec}}(X_{i}) = \left({\alpha_{\mathrm{r}}(X_{i})n_{\rm H} \left[\frac{f(X_{i+1})}{f(X_{i})}
-\frac{\alpha_{\mathrm{r}}(X_{i-1})}{\alpha_{\mathrm{r}}(X_{i})}\right]}\right)^{-1},
\end{math}\\
where $\alpha_{\mathrm{r}}(X_{i})$ is the recombination rate 
and $f(X_{i})$ is the fraction of element $X$ in ionization state $i$. 
If there is a change in column density between two
epochs, the time interval between the two measurements provides a lower limit on
the number density of the gas, leading to an upper limit on the distance of the
gas, using the definition for the ionization parameter in
Sect.~\ref{par:ion_par}. Conversely, a non-variation of the warm absorber as a function of the flux change
provides a lower limit on the distance. An ideal situation is when
the monitoring happens on different time scales. This allows to monitor both fast and slow 
flux changes and as a consequence to explore a wider range of possible densities. 
In practice such an ideal monitoring is difficult to perform. Brighter objects, which would
provide a high s/n, securing an accurate estimate of the parameters, do
not display much variability. Variations are of the order of
30\% using typical exposure times \citep[e.g.][]{2007A&A...461..121C,2010A&A...520A..36E}. 
The physical reason for this lack of dramatic variability 
is that probably they are larger systems, with larger black hole masses.
Therefore, unless dedicated monitoring is performed (e.g. Mrk~509, Kaastra et al, in prep.), a
comparison is done on data sets that are taken at different times, 
separated by months/years. Those time scales deliver only loose
constraints on the number density of the gas.\\ 
Only a handful of experiments were
successful in detecting significant variations of the warm absorber.  
For example, applying the relation for $\tau_{\mathrm{rec}}(X_{i})$ and for $\xi$ above, 
\citet{2005ApJ...622..842K} derived an upper limit of $r<6$\,pc for the
distance\footnote{Note that a non-variation of the warm absorber 
has been reported using a different data selection of the same observation \citep{2003ApJ...599..933N}.} of the gas in NGC~3783.
 A consistent limit of $r<25$\,pc has been derived for the same gas component 
from the analysis of the UV data \citep{2005ApJ...631..741G}. 
Using the variability method on NGC~3516, a distance estimate of 0.2\,pc 
has been obtained \citep[][]{2002ApJ...571..256N}.\\
Smaller black hole systems, like in the narrow line Seyfert~1 galaxies (NLS1),
show instead the desired large variation in flux on time scales of few ks. An intensively studied object in this respect is NGC~4051 \citep[e.g.][]{2007ApJ...659.1022K,2009A&A...496..107S}. 
Using different instruments, variability
studies point to densities which differ of a factor 20--40 for a same gas components. 
This mismatch underlines the challenge of this kind of analysis on complex spectra, 
as it relies on time-resolved spectroscopy on short time scales, which reduces the s/n considerably.\\

\subsection{Connection between gas emission and absorption}

Whether emission and absorption
in both UV and X-ray band are produced from the same gas it is not clear. The main issues are that the two
components do not vary on the same time scale and that the width of the
lines often do not match between absorbed and emitted lines. These facts point to a different location with respect to
the central source. A systematic
comparison between absorption and emission is often hampered by a low s/n and by the fact that 
in Seyfert~1 galaxies the continuum emission is predominant, favoring the detection of absorption, rather than emission, features. 
Absorption lines can therefore be compared only with a handful of
X-ray emission lines (typically the \ovii\ and \neix\ triplets, and \oviii) which are not
always detected. 
In the UV, the emission spectrum is dominated by the broad emission lines
whose width (2000-10,000\,\kms) cannot in general be reconciled with the narrower absorption
lines. Two extensively studied sources, NGC~4151
and NGC~5548 \citep{2008A&A...488...67D} recently underwent an historically broad-band low
flux. In those occasions, the UV broad lines were absent, revealing the presence of 
lines with smaller width (from an intermediate line region, ILR), consistently with that measured for the absorption
lines. Interpreting this width as due to Keplerian broadening, a distance
of 0.1 pc for both the emitting and absorbing gas has been derived in
NGC~4151 \citep{2007ApJ...659..250C}. This estimate could be readily applied to the
X-ray band absorbing gas, as this shared the same parameters as the 
UV absorbing/emitting  gas. 
The case of NGC~5548 did not deliver a similarly simple picture, as the ILR did not have any
detected absorption counterpart \citep{2009ApJ...698..281C}. 
In the X-rays a direct comparison between absorption and emission has been performed e.g. on NGC~3783, for which a
long \xmm\ exposure allowed such a study. The similarity of the widths of emission and absorption lines and the lack of
variability of the absorption features among other reasons implied a location of the gas in the narrow line region 
\citep{2003ApJ...598..232B}, in disagreement however with the estimates based on the absorption features.

\section{Do outflows contribute to feedback?}
Outflows are of potential crucial importance for the metal enrichment of
the surrounding medium (Sect.~\ref{par:intro}). The mass for unit time carried away from the system can be expressed as:\\
\begin{math}
\dot{M}_{out}= 8\pi \mu rN_{\rm H}m_p\Omega v,
\end{math}\\
where $\mu$ is the plasma mean molecular weight per proton ($\mu\sim 1.43$), $r$ is the radius, 
$N_{\rm H}$ is the gas column density, $m_p$ is the proton mass, $\Omega$ the global
covering factor and $v$ the outflow velocity. Collecting the values of $\dot{M}_{out}$ from classical Seyfert~1 galaxies with available 
distance estimates (see references in Table~1), we see that they  range between 0.01--0.06
$M_{\odot}$\,yr$^{-1}$. These values are often comparable with the mass accretion rate. In some outliers, $\dot{M}_{out}>10
\dot{M}_{acc}$ \citep[e.g.][]{2007ApJ...659..250C}.\\
The kinetic luminosity is simply given by\\
\begin{math}
L_{kin}=1/2\dot{M}_{out} v^2\,{\rm erg\ s}^{-1}.
\end{math} \\
The energetics is dependent on both observable that are easy to measure (e.g. $N_{\rm H}$, $v$) and on the
radius $r$, for which only in a handful of cases an estimate could be obtained. 
For a given radius, e.g. assuming an origin at the molecular torus distance 
\citep[$\sim$0.1--5\,pc in the sample presented in Table~\ref{t:list} of][]{2005A&A...431..111B}, then the highest column
density, and most of all, fastest outflows are likely to have an impact on the surrounding medium. For the Seyfert~1
considered here, the kinetic luminosity is in general modest: $L_{kin}<0.1$\%$L_{bol}$, 
where $L_{bol}$ is the bolometric
luminosity. Taking into account the sources where the $L_{kin}$ value was extracted, a range of
$10^{39.2-40.7}$\,erg\,s$^{-1}$ is
obtained. Considering an AGN life-time of $\sim4\times10^{8}$\,yr \citep[e.g.][]{2009A&A...500..749E}, the total energy
carried by the outflow is $10^{55.3-56.8}$\,erg. These values, keeping in mind the associated uncertainties, may be
comparable to the energy necessary to evaporate the interstellar environment out of the host 
galaxy \citep[$E\sim10^{57}$\,erg,][]{2010ApJ...710..360K}. An influence of the ionized flows on the intracluster medium and beyond is
difficult to assess. Simulations show indeed that only an energy of about $10^{60}$\,erg would make the outflows important at
intergalactic scale \cite[e.g.][]{2004ApJ...608...62S}.

\section{Conclusions}\label{par:conclusions}
Ionized outflows in Seyfert galaxies must play a role both in the self-sustenance of the active galactic nucleus and in
the chemical enrichment of the environment. Multiwavelength campaigns and dedicated monitoring 
on interesting objects are key in extracting the
most accurate parameters of the outflow. From that, also informations on the energetics of the outflow can be obtained.
Although our knowledge on warm absorbers has enormously increased in the last decade, critical parameters are still extremely
challenging to collect. The new HST instruments (COS, STIS) will allow easier density diagnostic studies, thanks to the
high throughput. In the X-ray future, IXO will allow large monitoring campaign of warm absorbers in AGN, with minimal exposure times. The new
technologies will thus allow us to quantify the impact of the outflows on the surrounding environment.

\begin{acknowledgements}
I wish to thank the organizers of the "High-resolution X-ray spectroscopy: past, present, and future" conference, 
held in Utrecht March 15-17 2010, for inviting me to present this review. 
I also thank Jerry Kriss, Jacobo Ebrero and the anonymous referee for their comments on this manuscript. 
\end{acknowledgements}


%
%

\end{document}